# A Privacy Glossary for Cloud Computing

Tian Wang
*Informatics Programs*
*University of Illinois*
Champaign, USA tianw7@illinois.edu

Masooda Bashir
*School of Information Sciences*
*University of Illinois*
Champaign, USA
mnb@illinois.edu

*Abstract*—Cloud computing is an evolving paradigm that is frequently changing the way humans share, store, and access their information in digital format. While cloud computing offers tremendous benefits (e.g., efficiency, flexibility, and reduced costs), it also brings both security and privacy challenges. Although cloud security has been extensively defined and developed, privacy protections in cloud environments are often described in abstract or vague language, which makes it difficult to interpret and implement. In this study, we propose an initial approach of developing a privacy glossary for cloud computing that provides a consistent and comprehensive set of terminologies for cloud privacy. We believe that this systematic and structured privacy glossary could serve as a first step towards implementing requirements for privacy protections in cloud computing, as well as providing more effective and consistent language in cloud privacy to researchers and professionals in the future.

*Keywords—Information Privacy, Cloud Computing, Privacy Glossary*

## I. INTRODUCTION

Nowadays, consideration of information privacy has become an essential element of all computing and online services. Information privacy refers not only to the confidentiality of personal information, but also to the protection of personal information and safeguarding of the collection, access, use, dissemination, and storage of personally identifiable information [1]. Meanwhile, the technology of cloud computing has shown advantages in delivering information technology infrastructure, components, and applications. The cloud environment could provide on-demand network access to configurable computing resources and minimize the need for management or service providers. However, with a large amount of personal and sensitive information being stored in the cloud environment, the development of cloud computing also poses a major threat to privacy because of its attractiveness as a target for cyber-attacks [2]. In the cloud environment, privacy issues include, but are not limited to what data should be collected, with whom the data will be shared, how the data is stored and transmitted, and who has access to it. Considering the complexity of privacy in cloud computing, solutions to information privacy issues in the cloud are increasingly needed to provide a safe, trustworthy cloud environment to promote users' confidence and economic development.

To address this need, there are several research and industry initiatives working towards better privacy protections in the cloud. However, one main challenge is the lack of a common language or set of terminologies (or "terms") on what constitutes privacy protections. In addition, it is often challenging to differentiate privacy and security protections. Although there exists an overlap between security and privacy in cloud computing, it is still important to distinguish privacy from security because each type of protection has a distinct purpose. Security often refers to the safeguarding of data and preventing or solving technical issues, while privacy in cloud computing is more complex to define since the concept itself is still being discussed without a consistent language. Not only is it quite vague to define privacy in cloud computing, but also the concept of privacy and relevant terminologies may have different descriptions in different regions or under different circumstances. For example, the General Data Protection Regulation (GDPR) takes a rights-based approach to privacy and guarantees certain protections to individuals in the EU, while the Federal Trade Commission (FTC) interprets privacy from a business perspective and focuses on protecting consumers' personal information through adjudication, policy initiatives, and consumer and business education.

Given the critical need of consistent understanding and communication on cloud privacy, one of the approaches is to develop a cloud privacy glossary that can be applied in various circumstances (e.g., risk assessment, certification evaluation, framework development) before conducting further study of information privacy in cloud computing. In general, a glossary could help individuals understand important terms related to specific fields. A previous study by Haas et al. [3] that developed a statistical glossary found that such a glossary could help to improve user's success and satisfaction in finding information by providing explanations in various presentations. Similarly, the cloud privacy glossary we propose in this study is expected to provide a much-needed common set of terms that would facilitate communication, the development of common privacy protection criteria, and the implementation of such protections at the global level. We believe it is a necessary and important step that can assist privacy professionals and researchers in tackling the complex concept of cloud privacy.

In this paper, we present our initial research into developing a privacy glossary for cloud computing based on a comprehensive list of privacy terms collected from a variety of sources. The proposed privacy glossary in this study aims to address the need for a reliable source of cloud privacy definitions, to offer a more comprehensive and systematic view of the privacy considerations in cloud computing, and to provide directions for future studies on cloud privacy.

## II. BACKGROUND

Although privacy has been addressed as one of the considerations when developing standards or frameworks under cloud computing environment, review of previous literatures showed that there hasn't been any consistent and comprehensive set of terms that specifically focus on cloud privacy yet. For example, an original work by Miller [4] mentioned building a taxonomy of big-data usage in cloud computing, while data privacy was only presented as part of the content instead of expanding it with details. Another paper by Sun et al. [5] discussed how organizations manage data security and privacy in cloud computing, but there was no clear distinction between security and privacy in their proposed framework. Previous research showed that privacy was often combined with security when discussing data protection in cloud environments. A previous study by

Idrissi et al. [6] mentioned the concepts of availability, confidentiality, and integrity in cloud computing, but it didn't clearly define privacy in the cloud. Another study by Shaikh and Samikumar [7] also only mentioned data privacy as one of the active areas of research and experimentations in cloud computing along with security when classifying relevant properties.

Currently, while different privacy-related glossaries have been developed, there is still no glossary specifically focusing on the aspect of cloud privacy. Some glossaries introduce privacy terms in general with no specification. The glossary of privacy terms collected by the International Association of Privacy Professionals (IAPP) provides privacy tools and information for professionals, but the main goal of this glossary is to help professionals develop their careers and to help organizations manage risks [8]. Some of the existing glossaries emphasize cybersecurity. One example is the NICCS Cybersecurity Glossary, which is a cybersecurity glossary by the National Initiative for Cybersecurity Careers & Studies that is intended to serve public and private sector cybersecurity communities. The glossary includes terms from other security-based documentations, such as CNSSI 4009 and NICE Framework [9]. The Computer Security Resource Center from NIST also developed the NIST CSRC glossary that consists of terms and definitions extracted verbatim from NIST's cybersecurity- and privacy-related Federal Information Processing Standards (FIPS), NIST Special Publications (SPs), and NIST Internal/Interagency Reports (IRs), as well as from Committee on National Security Systems (CNSS) Instruction [10]. In addition, there are some privacy glossaries that are designed for other aspects of privacy instead of cloud computing. For example, NIH Privacy Glossary, with the latest version of the glossary published in July 2017, is a glossary developed by Office of the Senior Official for Privacy in the National Institutes of Health [11].

Meanwhile, many of the existing privacy-related documents also include key concepts and definitions of privacy terms. The APEC Privacy Framework identified the core definitions and additional definitions to make clear the extent of coverage of the principles [12]. Similarly, Article 4 of the GDPR also introduces definitions of terms related to data protections [13]. Other developed privacy frameworks, such as ISO/IEC 29001, NIST Privacy Framework, and California Consumer Privacy Act (CCPA), also include specific sections defining privacy concepts. Considering different privacy frameworks and guidelines may define unique terms in their content, it is necessary to include concepts from these documents in order to develop a comprehensive cloud privacy glossary.

III. METHOD

The overall process of developing the proposed cloud privacy glossary is described in Figure 1. To develop the cloud privacy glossary, we first collected all the terms from 13 selected privacy-related sources (as listed in the figure). The retrieval of privacy terms was divided into two steps: first, we automatically retrieved the explicit privacy terms that clearly included privacy-related content in the definitions; second, we evaluated the remaining terms to see if any of them implied privacy protections in its definition. The initial list of terms, including both explicit and implicit privacy terms, was reviewed and discussed by researchers and professionals from the privacy fields. The list was updated based on the feedback received. To test the accuracy and comprehensiveness of the glossary, we also conducted an exploratory study by applying the explicit terms to various privacy-related documents to see the word frequency (i.e., how many terms from the glossary appear in the document, how many times each term appears). Considering the development of privacy protections in cloud computing, the cloud privacy glossary proposed in this study will be also updated in the future based on newly published content (e.g., updated privacy sources, newly developed privacy glossaries). The following sub-sections describe the details of each step to develop the glossary.

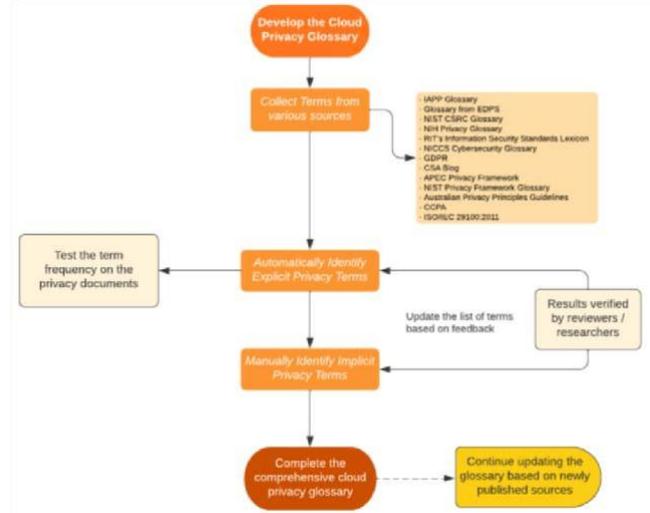

Fig. 1. Process of Developing the Cloud Privacy Glossary

*A. Collect privacy terms from a variety of privacy-related sources*

To select sources, we searched for all the publicly available documents that may include privacy-related terms or glossaries. The privacy terms were selected from two types of sources: 1) existing glossary or lexicons related to security and privacy, and 2) terms and definitions from privacy-related documents (e.g., standards, frameworks, and guidelines). For each term, the term name, its definition, and the original source it comes from, are recorded. After the initial process of term retrieval, a total of 6065 terms were collected. A list of all the sources selected for this study, as well as number of terms retrieved from each source, was shown in Table 1.

TABLE I. NUMBER OF TERMS RETRIEVED FROM EACH SOURCE

| Source | Number of Terms |
|---|---|
| NIST CSRC Glossary | 4653 |
| IAPP Glossary | 682 |
| NICCS Cybersecurity Glossary | 240 |
| NIH Privacy Glossary | 170 |
| Glossary from European Data Protection Supervisor | 78 |
| RIT's Information Security Standards Lexicon | 77 |
| NIST Privacy Framework Glossary | 38 |
| CSA (Blog) – 34 Cloud Security Terms | 34 |
| CCPA – 1798.140 Definitions | 28 |
| ISO/IEC 29100:2011 Information technology — Security techniques — Privacy framework | 27 |
| GDPR – Art. 4 Definitions | 26 |
| Australian Privacy Principles guidelines – Key concepts | 25 |
| APEC Privacy Framework – Core/Additional Definitions | 8 |

The 13 selected sources include:

- 6 glossaries: Current publicly available glossaries that are related to privacy protections, including NIST CSRC Glossary, NICCS Cybersecurity Glossary, IAPP glossary, NIH Privacy Glossary, NIST Privacy Framework Glossary, and Glossary from European Data Protection Supervisor. Since currently there is no widely applied set of terms that are specifically designed for cloud privacy protections, it is necessary to be more inclusive to ensure all the relevant terms were included during the process of term retrieval. While some of them (i.e., NIST CSRC glossary) were originally designed as a cybersecurity glossary, an initial review of the content found that some terms may also imply privacy protections.

- 1 lexicon: The Information Security Standards Lexicon developed by Rochester Institute of Technology. Like the cybersecurity glossaries, the terms used in the Information Security Standards may be applied to privacy protections as well.

- 6 privacy-related documents: Privacy-related documents (e.g., regulations, frameworks, standards, guidelines, principles) that identify a set of privacy terms in its content, including guidelines drawn from the CCPA, ISO/IEC, GDPR, CSA, APEC Privacy Framework, and Australian Privacy Principles. These documents include a separate section that contains a list of terms with definitions (for example, the key concepts from Australian Privacy Principles guidelines, or Core/Additional Definitions from APEC Privacy Framework). Although most of them only have a small number of terms, considering the original sources were specifically developed for privacy and data protections, it is important to include terms from these documents since they could be more directly related to privacy compared with the ones that were originally developed as cybersecurity terms.

Note that for the development of the cloud privacy glossary, we only selected sources that include general privacy terminologies with definitions. There are also some existing privacy-related documents (e.g., cloud certifications) with privacy requirements or controls that refer to the measures that provide information protection. These measures are not considered for this glossary since they are defined with specifications under certain context and are designed for practical cases, while the terms in the glossary are expected to provide a consistent language that can be applied to any circumstances in cloud privacy.

*B. Identify explicit and implicit privacy terms*

Since some of the selected sources focus on cybersecurity (e.g., NIST CSRC glossary is a cybersecurity-based glossary), and some privacy glossaries focuses on aspects other than those relevant to cloud computing (i.e., NIH Privacy Glossary focuses more on the health information aspect), the next step is to identify the privacy terms specifically related to cloud computing. Considering privacy is still an ambiguous concept in cloud computing, and many of the existing documentations combine privacy with security, we decided to identify terms related to cloud privacy with two categories: explicit privacy terms and implicit privacy terms.

**Explicit Privacy Terms**

Terms that explicitly refer to privacy are identified as the explicit privacy terms. The following steps were applied to retrieve explicit privacy terms:

1. Automatically retrieve the term if its definition includes any of these keywords: "privacy," "data protection," "personal information," "personal data," "confidentiality," "private," "data breach," and "PII."

2. The researcher manually reviewed the content of all the retrieved terms to ensure the definition explicitly relevant to cloud privacy, to eliminate the possibility of any opposite meaning. For example, if the definition of the term says, "this term is not developed as a privacy term," it would be manually excluded. Even though it includes the keyword "privacy", it shouldn't be considered as an explicit privacy term.

**Implicit Privacy Terms**

After identifying the explicit privacy terms, we continued reviewing the remaining terms and identified the ones that may imply privacy protections. One example of the implicit privacy terms is the term "data", which is defined as "a representation of information, including digital and nondigital formats, with the potential for adverse consequences for individuals when processed" in NIST Privacy Framework Glossary. Although it doesn't directly mention personal information in its definition, it explains that the use of data may lead to harmful outcomes to individuals, which implies the importance of personal data protection. Given the implied content from the definition of "data," we considered it as an implicit privacy term.

To identify the implicit privacy terms, all the remaining terms (excluding the explicit privacy terms) were reviewed by three researchers independently and marked with "Yes" (if the term implies any privacy protections in its definition) or "No" (if the term is not relevant to cloud privacy). After the individual process was completed, the terms reviewed by the three researchers were compared and then the three researchers reached consensus on if the term was implicitly relevant to privacy or not. The term was identified as implicit privacy term if two or more researchers marked it with "Yes".

*C. Remove overlapping terms*

During the process of collecting terms, we found that some terms may appear in more than one source. To remove the redundancy, we evaluated the definition of each term that is identified in multiple sources. For all the overlapping terms:

- If the term has exact same definition in different sources, the term and its definition will only be recorded once. For example, one of the explicit privacy terms, "privacy control," appears in NIH Privacy Glossary, NIST CSRC Glossary, and NIST Privacy Framework Glossary, with the same definition ("the administrative, technical, and physical safeguards employed within an agency to ensure compliance with applicable privacy requirements and manage privacy risk").

- If the term is defined differently in multiple sources, the definition from each source will be recorded to ensure that every aspect of the term definition is included. For example, one of the implicit privacy terms, "information sharing," appears in NICCS Cybersecurity Glossary and NIST CSRC Glossary, but it has different definitions (defined as "an exchange of data, information, and/or knowledge to manage risks or respond to incidents" in NICCS and defined as "the requirements for information sharing by an IT system with one or more other IT systems or applications, for information sharing to support multiple internal or external organizations, missions, or public programs" in NIST CSRC).

After removing redundancy, the initial version of the cloud privacy glossary includes 870 terms from 13 selected sources. Results and details are described in the following section.

IV. RESULTS

Table 2 summarizes the implicit and explicit privacy terms retrieved from each selected source, as well as the total number of privacy terms retrieved. Before removing the overlapped terms, a total of 958 privacy terms were collected, with 531 explicit privacy terms and 427 implicit privacy terms. It is no doubt that majority of the privacy terms are from the glossaries, considering the original glossaries include more terms than the privacy-related documents (most of the selected privacy-related documents only have one or two sections with few terms defined). After removing the overlapped privacy terms, our proposed cloud privacy glossary includes 870 unique privacy terms, with 478 explicit privacy terms and 392 implicit privacy terms.

TABLE II. NUMBER OF PRIVACY TERMS IN EACH SOURCE

| Source Name | Implicit Privacy Terms | Explicit Privacy Terms | Total Privacy Terms Retrieved |
|---|---|---|---|
| NIST CSRC Glossary | 148 | 136 | 284 |
| IAPP Glossary | 3 | 171 | 174 |
| NIH Privacy Glossary | 60 | 72 | 132 |
| NICCS Cybersecurity Glossary | 117 | 9 | 126 |
| Glossary from EDPS | 21 | 46 | 67 |
| RIT Standards Lexicon | 35 | 3 | 38 |
| NIST Privacy Framework Glossary | 15 | 16 | 31 |
| ISO/IEC 29100 | 3 | 23 | 26 |
| Australian Privacy Principles | 0 | 19 | 19 |
| GDPR Art.4 | 1 | 18 | 19 |
| CCPA 1798.140 | 8 | 11 | 19 |
| CSA Blog | 16 | 0 | 16 |
| APEC Core Definitions | 0 | 7 | 7 |

For most of the privacy frameworks, principles, and guidelines, the terms included in the cloud privacy glossary are explicit privacy terms. Given the fact that these documents are developed for privacy protections, it is reasonable that majority of the terms included are explicitly and directly relevant to privacy. In contrast, the privacy terms retrieved from glossaries that are developed for cybersecurity (i.e., NICCS, RIT Standards) are more likely to be implicit privacy terms. Since terms from these glossaries are originally identified as information security terms, it may only imply some privacy aspects in the content instead of directly referring to information privacy.

To have a better understanding of how inclusive each selected source is when referring to privacy terms, we calculated the percentage of privacy terms in each source (as shown in Figure 2). Most of the sources with a higher percentage of privacy terms were originally developed as privacy frameworks. For example, the ISO/IEC 29100 is a privacy framework with the goal to "provide a higher-level framework for securing Personally Identifiable Information with Information and Communication Technology systems," thus most of the terms included in ISO/IEC 29001 are expected to be relevant to cloud privacy protections. It is also understandable that the cybersecurity glossaries and lexicon (NICCS, RIT's standards, and NIST CSRC) may not include many privacy terms since the original source is from a security perspective and focuses more on the detailed techniques instead of protections of personal and sensitive data. Surprisingly, the IAPP glossary, which is designed as a privacy glossary, also has a lower percentage of privacy terms. After carefully reviewing the terms included in IAPP, we found that many of the terms are used to help professionals and organizations to understand the concept of privacy from a business or management perspective, instead of explaining how information privacy is protected under a cloud environment.

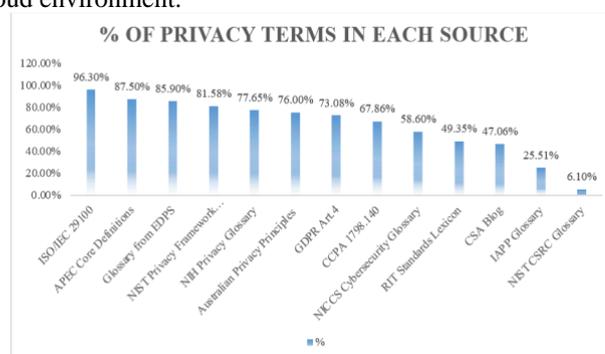

Fig. 2. Percentage of privacy terms in each source

V. DISCUSSION

While cloud computing continues to be a popular computing platform for all aspects of our lives, privacy protections in cloud computing have not received much attention. A recent literature review study [14] examined scholarly papers published from 2001 to 2020 that analyzed how cloud computing privacy has been discussed in the past twenty years. The authors of this review report that although cloud privacy has been widely mentioned and introduced, most of these papers didn't distinguish between the concepts of privacy and security; instead, many of the previous studies identified privacy as part of information security. Results from the proposed privacy glossary also show that it is often difficult to distinguish privacy terms from security in cloud computing, with about half of the terminologies (46.76%) originating in a cybersecurity glossary (i.e., NICCS Cybersecurity Glossary, NIST CSRC Glossary). Thus, it is apparent that we need to distinguish privacy terms from security terms and establish privacy-specific terminology if we are to be deliberate in providing privacy protections in cloud computing. This may mean that there will be some terms that are included in both classifications, but making this distinction is a critical step. It is also important to note, that though many of these privacy terminologies may have been originally specified as a security term, separating these terms, and focusing on the privacy aspects that are intended or meant within those terms are captured and preserved. For example, the term "data protection" could be used for both security and privacy. From a security point of view, the concept "data protection" refers to "the process of safeguarding important data from corruption, compromise or loss and providing the capability to restore the data [15]", which focuses more on the technologies that protect the data. On the other hand, as defined in IAPP, "data protection" from a privacy perspective refers to "the rules and safeguards applying under various laws and regulations to personal data about individuals that organizations collect, store, use and disclose", and it is different from data security because of its extension on devising and implementing policies for the fair use (IAPP). This type of distinction is an essential step towards building privacy protections that are not confused or lost in a security focused environment, thus, building a well-defined glossary specified for cloud privacy as the standard for researchers and practitioners is a critical step forward.

In addition, in our development of the cloud privacy glossary, we found that 88 terms appear in more than one source. While some of them have the exact same definition in different sources (one source directly used the definition from another source, or if two or more sources referred to the same origin), some of the terms have a unique definition in each source where it appears. The difference in definition and intended protection can be a source of ambiguity and loss of privacy protections. Therefore, constructing a glossary of terms that are focused on and intended for privacy protection would provide a common language and set of criteria that can be evaluated for privacy preservation that is not just a side effect of security protection. For example, the term "risk" appears in multiple sources and each source has a different definition.

- NIST Privacy Framework Glossary: "A measure of the extent to which an entity is threatened by a potential circumstance or event, and typically a function of: (i) the adverse impacts that would arise if the circumstance or event occurs; and (ii) the likelihood of occurrence."
- NICCS Cybersecurity Glossary: "The potential for an unwanted or adverse outcome resulting from an incident, event, or occurrence, as determined by the likelihood that a particular threat will exploit a particular vulnerability, with the associated consequences."
- NIH Privacy Glossary: "The net mission impact considering (1) the probability that a particular threat source will exercise (accidentally trigger or intentionally exploit) a particular information system vulnerability and (2) the resulting impact if this should occur."

Even though the three sources above are from different organizations (National Institute of Standards and Technology, National Initiative for Cybersecurity Careers & Studies, and National Institutes of Health), and they focus on different aspects, the definition of the term "risk" still has a similar meaning when comparing these sources. However, the differences in the definition may still cause confusion and require more effort for privacy practitioners to implement. Therefore, a glossary with consistent language could be helpful for professionals and researchers to have better communication in the future by increasing effectiveness and efficiency.

As privacy concerns become more important in the cloud environment, protecting personal data in the cloud represents a huge challenge, and it is important to clearly define the privacy terms in order to understand the concepts with a consistent language when discussing information privacy in cloud computing. Currently, there are only a few published research works that examine the adequacy of the existing cloud computing frameworks with respect to information privacy, and one of the reasons is the lack of comprehensive standards with a consistent language. The cloud privacy glossary proposed in this study could serve as an initial step towards building such consistent and comprehensive standard for cloud privacy, considering the different aspects covered by the selected sources, and the focus on the privacy aspect.

## VI. LIMITATIONS

In the initial research and process of building the cloud privacy glossary, we found that the definition of some privacy terms (especially terms that are implicitly related to privacy) might be ambiguous, and the identification of the implicit privacy terms was based on the researchers' understanding of privacy and the content of the definition. Therefore, further research is necessary to improve the accuracy and comprehensiveness by updating the glossary based on feedback received from the privacy community.

Also, the versions of selected sources for this study were the most recent versions that are publicly available online. However, it is possible that some of these documentations were updated after we developed the cloud privacy glossary, or that some of the relevant documents or glossaries are not publicly available. Therefore, we acknowledge that our results may be influenced by this situation, hence it is possible that our proposed glossary may miss some privacy terms from these sources.

## VII. CONCLUSION

In this research study, we proposed the development of a privacy glossary for cloud computing. While the process of developing was mainly qualitative and preliminary, this cloud privacy glossary represents our initial findings towards developing a common set of terms and a consistent language to interpret the complex concepts of privacy in cloud computing. Regardless of different geographical regions, cultures, and interpretations of privacy, the proposed glossary aims to take a step forward to maintain well-rounded privacy protections for cloud users, and to serve as the baseline for building or evaluating cloud privacy standards by bridging the knowledge gaps.

For the future works, to improve the accuracy of the proposed cloud privacy glossary, verification (e.g., using Natural Language Processing methods) is necessary to minimize the bias in the process of sources selections or the retrieval of terminologies. It is also important to review more relevant sources and receive feedback from the privacy community to ensure the privacy glossary is inclusive and comprehensive. Also, the privacy community is encouraged to suggest any additional terms for inclusion in the later versions of the proposed cloud privacy glossary. The ultimate goal of developing the cloud privacy glossary is to provide a consistent, effective way for researchers and organizations in different fields to better communicate on the topic of cloud privacy.

## ACKNOWLEDGMENT

This research study has been supported by Cisco Inc. We want to acknowledge and thank all of those who have contributed to this study.

## REFERENCES

[1] Di Giulio, C., Sprabery, R., Kamhoua, C., Kwiat, K., Campbell, R., & Bashir, M. N. (2017, May). IT security and privacy standards in comparison: improving FedRAMP authorization for cloud service providers. In 2017 17th IEEE/ACM International

Symposium on Cluster, Cloud and Grid Computing (CCGRID) (pp. 1090-1099). IEEE.

[2] Guilloteau, S., & Venkatesen, M. (2013). Privacy in Cloud ComputingITU-T Technology Watch Report March 2012.013

[3] Haas, S. W., Pattuelli, M. C., & Brown, R. T. (2003). Understanding statistical concepts and terms in context: The GovStat Ontology and the Statistical Interactive Glossary. *Proceedings of the American Society for Information Science and Technology, 40*(1), 193-199.

[4] Miller, H. E. (2013). Big-data in cloud computing: a taxonomy of risks.

[5] Sun, Y., Zhang, J., Xiong, Y., & Zhu, G. (2014). Data security and privacy in cloud computing. International Journal of Distributed Sensor Networks, 10(7), 190903.

[6] Idrissi, H. K., Kartit, A., & El Marraki, M. (2013, April). A taxonomy and survey of cloud computing. In 2013 National Security Days (JNS3) (pp. 1-5). IEEE.

[7] Shaikh, R., & Sasikumar, M. (2015). Data Classification for achieving Security in cloud computing. Procedia computer science, 45, 493-498.

[8] Glossary of Privacy Terms. IAPP. Retrieved from (https://iapp.org/resources/glossary/).

[9] NICCS Cybersecurity Glossary. Retrieved from (https://niccs.cisa.gov/about-niccs/cybersecurity-glossary#).

[10] NIST CSRC Glossary. Retrieved from (https://csrc.nist.gov/glossary).

[11] NIH Privacy Glossary. Retrieved from (https://oma.od.nih.gov/DMS/Documents/Privacy/NIH%20Privacy%20Glossary%202017.pdf).

[12] APEC Privacy Framework – Core Definitions/Additional Definitions. Retrieved from (https://www.apec.org/-/media/APEC/Publications/2005/12/APEC-PrivacyFramework/05_ecsg_privacyframewk.pdf).

[13] Art. 4 Definitions. GDPR. Retrieved from (https://gdpr-info.eu/art-4gdpr/).

[14] Sharma, T., Wang, T., Giulio, C. D., & Bashir, M. (2020, October). Towards Inclusive Privacy Protections in the Cloud. In International Conference on Applied Cryptography and Network Security (pp. 337359). Springer, Cham.

[15] "What is Data Protection?" (n.d.) The Storage Networking Industry Association (SNIA). Retrieved from (https://www.snia.org/education/what-is-data-protection).